\documentclass{PoS}
\usepackage[authoryear,square]{natbib}
\bibpunct{(}{)}{;}{a}{}{,}

\title{Thermal radio emission from novae \& symbiotics with the Square Kilometre Array}

\ShortTitle{Novae \& symbiotic stars in the SKA era}

\author{
Tim O'Brien$^1$, 
\speaker{Michael Rupen}$^2$, 
Laura Chomiuk$^3$, 
Valerio Ribeiro$^{4,5}$, 
Michael Bode$^6$
Jennifer Sokoloski$^7$, 
Patrick Woudt$^4$
\\
$^1$University of Manchester;
$^2$NRC/CNRC-Canada; 
$^3$Michigan State University;
$^4$University of Cape Town;
$^5$Radboud University;
$^6$Liverpool John Moores University;
$^7$Columbia University.

E-mail: \email{tim.obrien@manchester.ac.uk}
}

\abstract{Thermal radio emission is a fundamental tracer of outflows from stellar systems. Novae and symbiotic stars are interacting binary systems incorporating accretion and nuclear burning on white dwarfs. They share several distinct observational features during their outbursts: higher flux densities at higher radio frequencies, variability on a range of (unpredictable) timescales, and locations largely confined to the Galactic plane. Physical insights 
drawn from free-free radiation are distinct and complementary to the more commonly sought after 
radio synchrotron emission, and high-quality observations of thermal processes draw on different SKA capabilities.

The thermal radio emission of novae during outburst enables us to derive fundamental quantities such as the
ejected mass, kinetic energy, and density profile of the ejecta. Recent observations with newly-upgraded
facilities such as the VLA and e-MERLIN are just beginning to reveal the incredibly complex
processes of mass ejection in novae (ejections appear to often proceed in multiple phases and
over prolonged timescales). Symbiotic stars can also exhibit outbursts, which are sometimes accompanied by the expulsion of material in jets. However, unlike novae, the long-term thermal radio emission of symbiotics originates in the wind of
the giant secondary star, which is irradiated by the hot white dwarf. The effect of the white dwarf on
the giant's wind is strongly time variable, and the physical mechanism driving these variations
remains a mystery (possibilities include accretion instabilities and time-variable nuclear burning
on the white dwarf's surface).

The exquisite sensitivity of SKA1 will enable us to survey novae throughout the Galaxy, unveiling statistically complete populations. With SKA2 it will be possible to carry out similar studies in the Magellanic Clouds. This will enable high-quality tests of the theory behind accretion and mass loss from accreting white dwarfs, with significant implications for determining their possible role as Type Ia supernova progenitors. Observations with SKA1-MID in particular, over a broad range of frequencies, but with emphasis on the higher frequencies, will provide an unparalleled view of the physical processes driving mass ejection and resulting in the diversity of novae, whilst also determining the accretion processes and rates in symbiotic stars.
}

\FullConference{
Advancing Astrophysics with the Square Kilometre Array\\
June 8-13, 2014\\
Giardini Naxos, Sicily, Italy}

\begin{document}

\section{Introduction}

Novae and symbiotic stars are interacting binary star systems in which a hot white dwarf orbits a companion main-sequence or red giant star. Material is accreted onto the white dwarf, via Roche-lobe overflow (typically in the case of novae) or directly from the wind of the companion. They are excellent laboratories for studying a number of very important astrophysical phenomena, for example: binary evolution; accretion; ionization of circumstellar material; mass ejection; jet formation; thermonuclear burning and explosions. Those with massive white dwarfs may eventually lead to Type Ia supernovae.   

Novae are the most common stellar thermonuclear explosions \citep{2008clno.book.....B}. They result from a thermonuclear runaway (TNR) in material accreted from a main sequence or red giant star onto the surface of a white dwarf. In the explosion, $10^{-8}-10^{-3}$ solar masses are ejected at speeds ranging from a few hundred to several thousand ${\rm km\ s}^{-1}$. The white dwarf is not destroyed so accretion begins again, leading to repeated explosions with recurrence periods ranging from a year (for recurrent novae in which more than one outburst has been observed, \citealt{2014ApJ...786...61T}) to millions of years (in the case of so-called classical novae, \citealt{2005ApJ...623..398Y}). If all the mass accreted since the previous outburst is not ejected, then novae with massive CO white dwarfs will evolve towards explosion as a Type Ia supernova. 

Radio observations are particularly useful in studies of novae as they provide a direct view of the ejected mass by tracking the free-free photosphere as it moves back through the outflow over months to years \citep{2008CNe.SeaquistBode, 2012BASI...40..293R}. The mass ejected can be estimated using standard models assuming free-free emission from instantaneous ejection of spherical shells \citep[e.g.][]{1979AJ.....84.1619H, 1977ApJ...217..781S, 1980AJ.....85..283S}. 
Radio imaging can of course also resolve the ejecta throughout the outburst, aiding distance estimates and determinations of ejection geometry \citep{1988Natur.335..235T, 1996MNRAS.279..249E}.

Recent observations of novae (see below) have resulted in discoveries which present challenges to the orthodox interpretation: nova ejecta are often aspherical, with some even showing evidence for jets \citep{2008ApJ...688..559R, 2008ApJ...685L.137S}; light curves indicate that there is frequently more than one major ejection \citep{2011ApJ...739L...6K, 2014ApJ...785...78N}; some novae show evidence for non-thermal radio emission \citep{2014ApJ...788..130C}; observational mass estimates tend to exceed predictions from models of the TNR by an order of magnitude \citep{2012BASI...40..293R}; and several novae have now been detected as high-energy gamma-ray sources, suggesting large-scale particle acceleration in shocks \citep{2014Sci...345..554A}. 

Symbiotic stars are wide binary systems where mass is accreted onto the surface of a white dwarf from a red giant star. These systems have high mass loss rates in the red giant wind, of order 10$^{-7}$ to 10$^{-6}$ solar masses per year 
\citep{1984ApJ...284..202S,1990ApJ...349..313S}. Unlike typical novae, thermal radio emission arises in symbiotics primarily from this wind ionised by the white dwarf. Its study therefore allows us to explore the circumstellar environment of these systems, the mass-loss history of the central system, and the luminosity of the white dwarf. Symbiotic stars undergo occasional outbursts in which the optical brightness increases and sometimes even thermonuclear-powered nova events (see below).
 During these outbursts, some symbiotic stars have shown evidence for collimated jets \citep{1995MNRAS.272..843D, 2001MNRAS.326..781C, 2010ApJ...710L.132K}. Mass accretion and nuclear burning on the white dwarf leads to the suggestion that they could be one class of progenitors of Type Ia supernovae \citep{1992ApJ...397L..87M, 2010ApJ...719..474D}.

Progress in understanding the nova phenomenon and its wider links to binary evolution is challenged by the heterogeneous nature of the population, particularly as revealed in recent observations with the VLA \citep{2012BASI...40..293R}. 
SKA will enable a statistically complete survey of novae and symbiotic stars in our Milky Way, providing multi-frequency light curves and resolved images of individual objects which will help us answer several major outstanding questions, including:\\
\hspace*{0.5cm} -- How much mass is ejected in nova outbursts and what system parameters determine the ejecta mass?\\
\hspace*{0.5cm} -- What is the geometry of mass ejection and what is its relation to the properties of the binary system?\\
\hspace*{0.5cm} -- What are the physical mechanisms driving mass loss during the nova outburst?\\
\hspace*{0.5cm} -- What is the density and distribution of circumbinary material?\\
\hspace*{0.5cm} -- What is the true galactic population of novae and symbiotic stars?\\
\hspace*{0.5cm} -- How do symbiotic stars transfer mass? Is it via a wind, Roche lobe overflow or a combination of both?\\
\hspace*{0.5cm} -- What proportion of symbiotic systems have ``jets"?\\
\hspace*{0.5cm} -- How do novae and symbiotic stars relate to the progenitors of Type Ia supernovae?

\section{Recent observations}

\subsection{Novae}

The 2006 outburst of the recurrent nova RS Ophiuchi was the target of an extensive multi-frequency campaign of observations which included: the first resolution of the expanding shock wave from a nova explosion using the VLBA, EVN \& MERLIN \citep{2006Natur.442..279O}; detection of X-ray emission from the expanding shock wave using RXTE \& Swift \citep{2006Natur.442..276S, 2006ApJ...652..629B}; the discovery of jet-like radio emission using the VLA \& VLBA \citep{2008ApJ...685L.137S, 2008ApJ...688..559R, 2009MNRAS.395.1533E}; and extensive coverage of the super-soft phase of X-ray emission using Swift \citep{2011ApJ...727..124O}.

Following the successful campaign on RS Oph, two large collaborations were formed: Swift-nova-cv uses Swift to study super-soft and shock X-ray emission from novae, whilst the E-Nova team (many of whom are also members of Swift-nova-cv) concentrates on exploiting new radio facilities such as the upgraded VLA and e-MERLIN to monitor novae in outburst. Both projects are resulting in far more extensive observational databases. Several examples of recent results are summarised below before we go on to discuss the future role of SKA.
 
T Pyx is a recurrent nova whose most recent outburst was in 2011. The VLA radio light curve showed several surprises \citep{2014ApJ...785...78N}. No detections for the first two months after the optical outburst were followed by a dramatic rise after day 65, and surprisingly high peak luminosities, requiring an ejection of significantly more mass than expected from the high-mass white dwarfs thought to host recurrent novae. To explain the rapid rise, this needs to be combined with significant cooling/reheating and/or delayed ejection (a hypothesis supported by the \emph{Swift} X-ray light curve, \citealt{2014ApJ...788..130C}). 

\begin{figure}
\includegraphics[width=.48\textwidth]{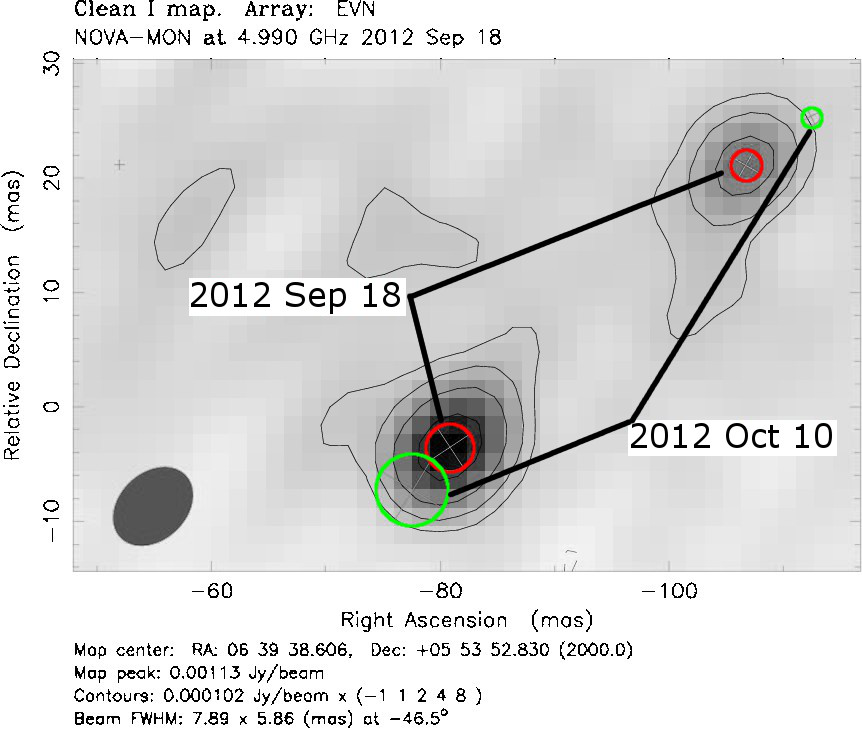}
\hfill
\includegraphics[width=.48\textwidth]{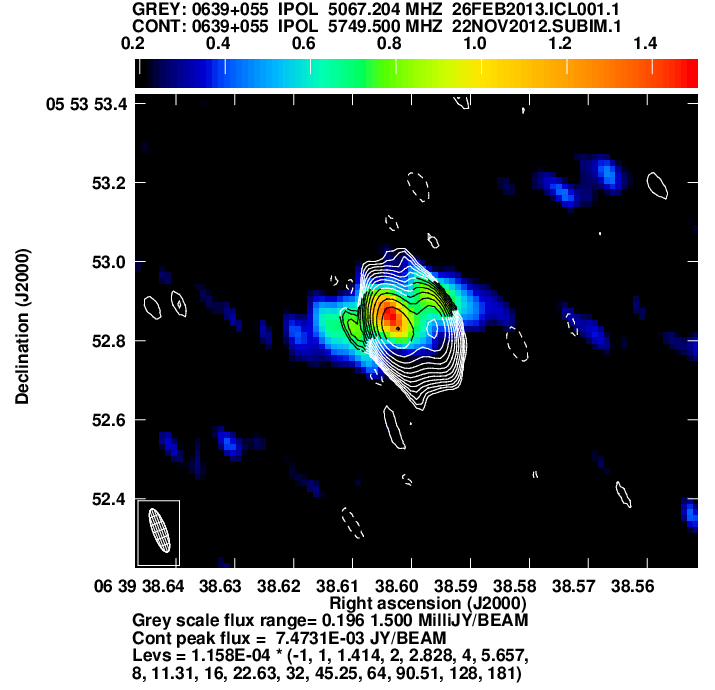}
\caption{Left: EVN image of V959 Mon on 2012 Sep 18 (91 days after \emph{Fermi} discovery; 5.8 GHz). Components in this image are labelled by red circles, with the positions of components 22 days later (Oct 10; day 113) labelled with green circles \citep[see][]{2014Natur.514..339C}. Right: e-MERLIN images on Nov 23 (day 157; 5.8 GHz) in contours and 2013 Feb 26 (day 252; 5.1 GHz) in colour scale (Healy et al. in prep).}
\label{figV959}
\end{figure}

V959 Mon is a classical nova which went into outburst in 2012. This was the third nova to be detected in GeV gamma-rays \citep{2014Sci...345..554A}, this time detected by \emph{Fermi} before it was identified as a nova due to its proximity to the Sun during outburst. VLA detected V959 Mon within a week or so of outburst, but three months later the source brightened significantly. Target-of-opportunity observations were obtained with e-EVN, e-MERLIN and VLBA \citep{2012ATel.4408....1O, 2014Natur.514..339C}, revealing two compact VLBI components moving away from each other and a larger-scale thermal bipolar structure (Figure \ref{figV959}). 

Since V959 Mon has a main sequence companion \citep{2013MNRAS.435..771M}, and is therefore not an embedded nova like RS Oph (i.e. with no dense pre-existing circumstellar medium), the gamma-ray and synchrotron radio emission must arise as a result of internal shocks between fast and slow-moving material ejected in the outburst. Radio imaging of V959 Mon \citep{2014Natur.514..339C} shows the structure evolve from an east-west structure to a larger north-south structure.  One possibility is that interaction of the expanding nova ejecta with the binary in a common-envelope phase expels material at different rates. This leads to the shock interactions visible as non-thermal emission with VLBI, and sculpts the expanding ejected shell, which is visible in thermal radio emission. 

\subsection{Symbiotic stars}

To date the largest radio survey of symbiotic stars is that from \citet{1993ApJ...410..260S}, detecting 50 stars (out of 99 observed) with a sensitivity of 110~$\mu$Jy. These were single-epoch observations, so much information was missed. For example, \citet{2004MNRAS.347..430B} observed Z~And over several epochs and frequencies, revealing variable radio emission and jet ejection.  Jet-like, collimated, or bipolar outflows have also been observed at radio frequencies in CH~Cyg \citep{2001MNRAS.326..781C} and R~Aqr \citep{1995MNRAS.272..843D}. In CH~Cyg, there is evidence for an episodic precessing jet (visible in the optical, X-ray and radio) interacting with the ambient medium \citep{2010ApJ...710L.132K}. 
The timescales, amplitude, and mechanism of radio variability remain poorly explored in symbiotic stars, but certainly contains information about the accretion, nuclear burning, and outflows that power these systems.
 
Perhaps even more fundamentally, recent detections of strong hard X-ray emission (E > 2 keV) from symbiotic stars that have weak signatures of mass transfer in their optical spectra (e.g. \citealt{2007ApJS..170..175B, 2007ApJ...660..651N, 2007ApJ...671..741L, 2009ApJ...701.1992K, 2013A&A...559A...6L}) suggest that past surveys of symbiotic stars have been subject to severe selection biases.  Since quasi-steady nuclear shell burning releases approximately 50 times more energy per nucleon than accretion alone, symbiotic white dwarfs powered by accretion alone are typically much less luminous -- with much lower fluxes of ionizing photons, weaker optical emission lines, and lower radio fluxes -- than symbiotic white dwarfs with shell burning.  Thus the strength of optical emission lines, which have been considered the defining feature of symbiotic stars, is not necessarily a good proxy for the rate of mass transfer.  Since most known symbiotic stars have been identified because of their strong optical emission lines, the Galactic population of symbiotic stars might actually be much larger than previously appreciated. Sensitive radio observations with SKA1 could both characterize the radio properties of accretion-powered symbiotic stars and also provide a new measure of the true Galactic population of these important interacting binaries.

\begin{figure}
\includegraphics[width=.4\textwidth]{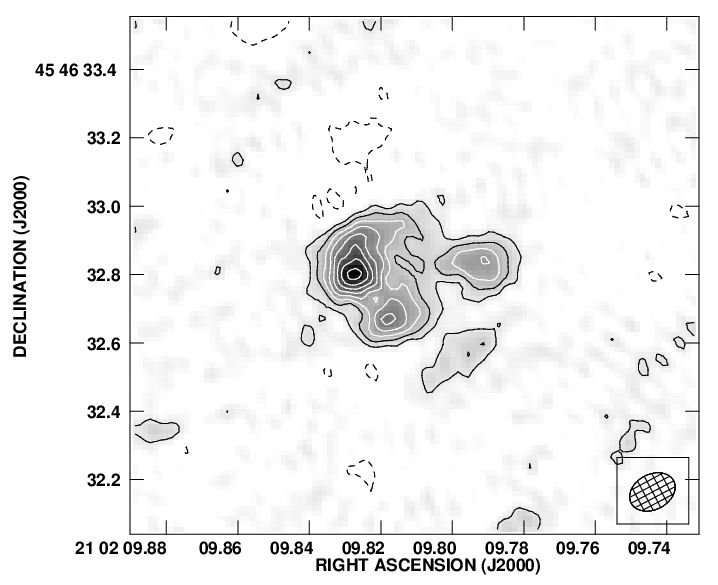}
\hfill
\includegraphics[width=.58\textwidth]{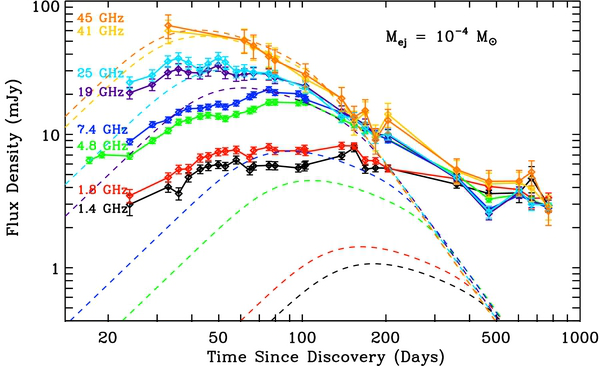}
\caption{Left: 5 GHz MERLIN image of V407 Cyg, 11 days after outburst, showing extended emission from the Mira giant wind. Right: VLA light curve showing how standard models of thermal radio emission from novae (dashed lines) fail to fit observations (points and solid lines) \citep{2012ApJ...761..173C}.}
\label{figV407}
\end{figure}

The symbiotic system V407 Cyg, composed of a white dwarf and a Mira giant, underwent a nova outburst in 2010 and was the first nova to be detected in gamma rays \citep{2010Sci...329..817A}. Radio observations were made with the VLA, MERLIN and EVN. MERLIN resolved the emission region within the first few weeks after outburst (Fig. \ref{figV407}), and showed that it was too extended to be from material ejected in the outburst and rather appears to be flash-ionized circumstellar medium. As the later VLA light curve showed (Fig. 2, \citealt{2012ApJ...761..173C}), the subsequent radio emission appears to originate from the extended wind of the Mira companion, increasing in luminosity as the wind is ionized, and fading as the wind is heated from within by the expanding nova shock wave \citep{2012ApJ...748...43N}---the first time such a phenomenon has been observed.  

\section{The role of SKA in studies of novae and symbiotic stars}
 
The first novae to be detected in the radio were HR Del and FH Ser in 1970 \citep{1970ApJ...162L...1H}. But as \citet{2008CNe.SeaquistBode} pointed out, as of 2008 only 17 classical novae had been detected in outburst and of these, only four had so-called first class data with multiple epochs and observing frequencies. Hence, novae with high-quality radio observations are too dissimilar and too few in number to know which behaviours are odd and which common, let alone to connect the mass ejection histories to the details of the host system. Novae are famously heterogeneous with a wide range of behaviour seen in the optical, including light curves whose decline times range from 4 to 900 days. Factor in the variation in circumstellar environments (e.g. whether the nova is embedded in a dense wind from the secondary) and the viewing angle and it becomes clear that a fuller understanding will require study of a statistically significant sample. 

Excellent progress is being made with current VLA monitoring programmes led by authors on this chapter, but even here we are limited to only a small subset of new novae, typically up to four new targets each year out of around ten discovered optically (current rate). Note, it is estimated that there are $\sim35$ nova outbursts in the Milky Way each year \citep{1997ApJ...487..226S} with the majority passing unnoticed in the optical, presumably because of distance, short duration at peak, high extinction, or proximity to the Sun at outburst. Even the outbursts of naked eye novae have gone unnoticed for weeks to months in recent times \citep{2010ApJ...724..480H}. In the era of LSST and SKA, this will change.

In a very simple model of thermal radio emission from expanding ejecta \citep{2008CNe.SeaquistBode}, the emission is initially optically thick and the flux rises as the solid angle increases. As its density decreases it becomes optically thin, first at higher frequencies, and the nova fades. This simple model (for a shell with mass $M_{ej}$, velocity $V_{ej}$ and electron temperature, assumed isothermal, $T_{e}$) leads to estimates for the maximum flux density $f_{max}$ at time $t_{max}$ when the angular diameter is $\theta_{max}$ :
\begin{equation}
f_{max} \approx 6.7 \left ( \frac{\nu}{\textrm{GHz}} \right )^{1.16} \left ( \frac{T_{{e}}}{10^4\ \textrm{K}} \right )^{0.46} \left ( \frac{M_{{ej}}}{10^{-4}\ \textrm{M}_{\odot}} \right )^{0.80} \left ( \frac{D}{\textrm{kpc}} \right )^{-2}\quad \textrm{mJy\ ;} 
\end{equation}
\begin{equation}
t_{max} \approx 2.6 \left ( \frac{\nu}{\textrm{GHz}} \right )^{-0.42} \left ( \frac{T_{{e}}}{10^4\ \textrm{K}} \right )^{-0.27} \left ( \frac{M_{{ej}}}{10^{-4}\ \textrm{M}_{\odot}} \right )^{0.40} \left ( \frac{V_{{ej}}}{10^3\ \textrm{km\ s}^{-1}} \right )^{-1} \quad \textrm{years\ ;}
\end{equation}
\begin{equation}
\theta_{max} \approx 1.1 \left ( \frac{\nu}{\textrm{GHz}} \right )^{-0.42} \left ( \frac{T_{{e}}}{10^4\ \textrm{K}} \right )^{-0.27} \left ( \frac{M_{{ej}}}{10^{-4}\ \textrm{M}_{\odot}} \right )^{0.40} \left ( \frac{D}{\textrm{kpc}} \right )^{-1} \quad \textrm{arcsec\ .}
\end{equation}

Slightly more complex models include a density gradient in a thick shell of ejecta, while continuing to assume that the ejecta are isothermal and spherically symmetric. These models describe mass ejection in either an impulsive, homologous explosion with a range of velocities (also known as a `Hubble flow'; \citealt{1977ApJ...217..781S}) or a prolonged wind \citep{1983MNRAS.202.1149K}. As with the simple model, flux density during the optically thick and thin phases is time dependent and is proportional to $\nu^2t^2$ and $\nu^{-0.1}t^{-3}$, respectively.

\begin{figure}
\centering
\includegraphics[width=0.8\textwidth]{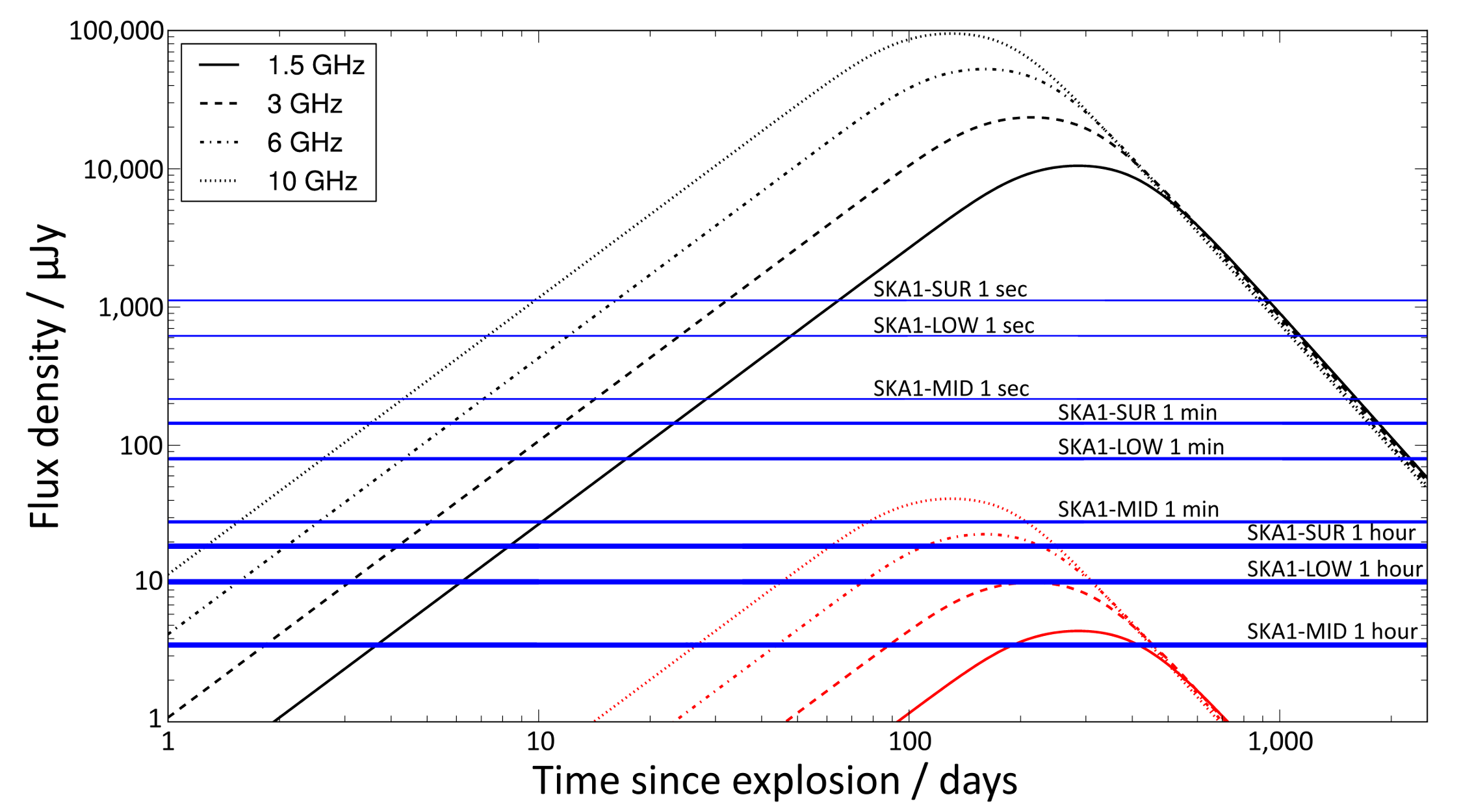}
\caption{Synthetic light curves at several radio frequencies for a nova with $M_{\rm ej} =~1~\times~10^{-4}$~M$_\odot$, maximum ejecta velocity $V_{{max}}$~=~3,000~km~s$^{-1}$, $T_{\rm e}$ = 17,000 K, and the ratio of the inner to outer radius of the shell of 0.25. The upper set of curves (black) are for a nova at a distance of 1~kpc whilst the lower set (red) are for the distance of the LMC. The horizontal blue lines are the 5-$\sigma$ sensitivity limits for SKA1 observations at 1 sec, 1 min, and 1 hour integrations.}
\label{sphere}
\end{figure}

In Figure~\ref{sphere} we show `Hubble flow' model light curves at various frequencies along with SKA1-SUR, -LOW, and -MID 5-$\sigma$ sensitivities at 1 second, 1 minute and 1 hour integrations \citep{Dewdney2013, Braun2013}. What is immediately evident from this figure is that we can detect nearby novae with very short integration times within a few weeks of the outburst. One minute integrations would allow us to characterise much of the light curve for novae within about 10 kpc. In addition, SKA1 will enable detection of novae out to 50 kpc (i.e., the Magellanic Clouds) with 1 hour integrations. It is clear from this figure that the most natural match to nova science is SKA1-MID at the higher frequency bands.

We will be able to carry out a complete survey of novae throughout the Galaxy, exploring the entire range of ejecta masses and timescales. SKA1 will probe not only the nearest and brightest novae, but the full population---i.e., $\sim$35 novae per year (compare with our current {\it total} of $\sim$20 with high-quality data). Radio surveys with SKA1 will see through the dust of the Galactic plane to reveal well-defined, complete samples of Milky Way novae, enabling detailed modeling of novae as an evolving population. Searches for novae can piggy back on wide-field surveys, with a focus on the Galactic plane ($|b| < 10^{\circ}$; if the Galactic plane were observed every few weeks, we could in principle detect every nova in the Galaxy visible from SKA latitudes).  
With SKA1, it  will be possible to map out complex ejection histories for complete samples, comparing radio evolution for different speed classes, sources with and without narrow optical lines, sources with and without hard X-ray and gamma-ray emission, etc. By using complementary observations at other wavebands we will be able to tie ejection mass, velocity, and evolution to properties of the white dwarf (e.g. composition, mass) and binary system (e.g. orbital period, companion mass, companion wind). 

SKA1 will enable, for the first time, detection of radio emission from novae in the Magellanic Clouds. The SKA1-MID 1-hour sensitivity of a few $\mu$Jy rms at 50 kpc is equivalent to several 10's of $\mu$Jy rms at the Galactic Centre, or around 0.1~mJy at a few kpc (Figure \ref{sphere}). By observing the LMC and SMC for 1 hour per pointing every month, we should easily detect most Magellanic Cloud novae with radio light curves comparable to those studied in the Milky Way today, allowing us to compare nova populations as a function of host metallicity.

In addition to the significant advantages of full population surveys, SKA1 will provide completely new information on individual novae. Sensitive early observations will allow us to look for rapidly-recombining emission from the surrounding circumbinary emission (as in the one example so far, V407 Cyg; Fig.~2). We will be able to detect sources when they have very small angular diameters and/or very low ejected mass, thereby determining exactly when the first mass is ejected and relating mass ejection to models of the TNR (e.g., \citealt{2014MNRAS.437.1962H}). This early faint radio emission will be detected when the hard X-ray and gamma-ray emission is also being detected, probing the mechanism for producing strong shocks in novae.

Excellent light curves will be obtained with very little commitment of observing time so we can increase cadence to daily or weekly observations and carry on observations for years. Such unprecedented time coverage will enable us to see exactly what is happening without theoretical bias (cf. the early rapid rise in RS Oph seen by MERLIN and the multi-peaked behaviour seen in recent VLA observations of several novae; \citealt{2009MNRAS.395.1533E, 2011ApJ...739L...6K, 2014ApJ...785...78N}).

Detection of radio recombination lines will allow us to map out the full 3D structure of novae. At the very least we could measure the strength and velocity structure of the lines. Simple estimates (e.g. following \citealt{CondonRansom}, \citealt{2002ASSL..282.....G}), suggest that the lines could be up to around 0.5\% of the continuum at the higher frequency bands. If the nova is spatially resolved while still optically thick, then 3D imaging with direct density and filling factor measures is possible. Sources at tens of mJy should provide measurements of HI and OH absorption, enabling distance constraints and direct measurements of the evolving neutral and molecular circumbinary medium.

The $\sim$100~km baselines at 10 GHz for SKA1-MID provides 60~mas resolution, better than VLA in A configuration (albeit with few baselines) and similar to e-MERLIN. This {\it uv} coverage will provide superb imaging 12 months a year at the higher frequencies (rather than only one quarter of the time with VLA).

Imaging with SKA1 will enable consistent observation of the evolving ejecta structure and direct measurement of the ejecta brightness temperature.
The high sensitivity of SKA1 means we can robustly compare radio images with, for example, JWST images, probing the nova structure with high time and angular resolution across the electromagnetic spectrum, and revealing the properties of the ejecta in unprecedented detail.

Currently we only detect one old nova via shock interaction with circumstellar material (GK Per; \citealt{1989ApJ...344..805S}). This interaction in GK~Per has enabled us to piece together the binary's history and probe the common envelope ejection that created the current binary system. We will be able to explore GK Per itself in extreme detail, with high sensitivity and at 60~mas resolution, and we will also be able to search for emission from other old novae to probe their circumstellar environments.

For symbiotic stars, we will be able to determine why some systems experience quasi-steady shell burning, and the pervasiveness of this phenomenon.  Nuclear shell burning could be the result of continuous accretion onto the white dwarf at a rate of a few times $10^{-8}-10^{-7}\ \textrm{M}_{\odot}$ yr$^{-1}$, as in the persistent supersoft X-ray sources (e.g. \citealt{1992A&A...262...97V}). The range of accretion rates required to produce quasi-steady shell burning, however, is quite narrow \citep{1982ApJ...257..767F}. Given the diversity of symbiotic systems with shell burning (orbital periods range from hundreds of days to decades), it is a challenge to explain how they could all accrete within the required narrow range of rates. Alternatively, shell burning could be residual nuclear burning from a previous nova explosion. In either case, ubiquitous steady burning would mean that most symbiotics can avoid full-scale nova explosions in which the entire accreted envelope is ejected, allowing symbiotic white dwarfs to accumulate material efficiently. Such efficient mass accumulation would increase the likelihood that they can gain enough material to explode as Type Ia supernovae. Since we expect accretion-powered symbiotics to produce $\sim$100 times less radio emission than symbiotics powered by nuclear shell burning, a sensitive radio survey like that which can be provided by SKA1 is required to find accretion-powered symbiotics and reveal the prevalence of shell burning.
	
With the reduced sensitivity of 50\% during the early phase of SKA1-MID, we will still be able to achieve a good part of the science case. However, we emphasise that Band 5 capabilities are crucial for studies of novae and symbiotic stars. 

\section{Optimising SKA for Thermal Transients}

Here, we address aspects of  SKA1 design which will enable us to meet our science goals described above.

Most importantly, studies of novae and symbiotics require a high frequency band (as high as possible). The most efficient band for observing optically-thick thermal emission, by far, will be Band 5. In addition, the radio spectral energy distribution holds critical information, so  quasi-simultaneous multi-band observations are key.  Such observations may lend themselves to the use of sub-arrays or multiple beams.

Observations of thermal transients requires efficient, short (seconds to minutes), and frequent (daily or weekly) observations over the lifetime of an outburst lasting months to years. Observations with the VLA have illustrated how unpredictable and poorly understood the radio light curves of novae are on timescales of days to weeks \citep{2012BASI...40..293R}, so it is important to be able to trigger on short timescales (within hours), particularly given that SKA will open up a new parameter space of low flux density observations early in the outburst. In addition, longer sensitive observations will enable study of short timescale variability, an aspect of accreting white dwarfs which remains completely unexplored in the radio (but may be expected, as such behaviour is seen at other wavebands; e.g. X-ray and optical).

To probe the complex evolving structures of novae and symbiotics, the angular resolution of SKA1 needs to be high, demanding Band 5 and antennas spread over several hundred kilometres. In addition, a number of novae have been imaged now with VLBI (e.g. RS Oph and V959 Mon) so SKA-VLBI will be key to imaging and tracking non-thermal components of emission resulting from the shocks now known to be present in many novae. A combination of lower-resolution imaging of the thermal ejecta and VLBI imaging of non-thermal components will be crucial for developing models to explain the evolution and ultimate fate of novae, helping us answer questions regarding the ejected mass, its geometry, and the role of shock interactions.

Study of radio recombination lines and disentangling multiple spectral components demands wide bandwidths with well-behaved bandpasses. In addition, a consistent flux density scale is needed to allow assessment of the reality of small changes in flux density caused by, for example, variations in ejection properties or other evolution of these dynamic sources.

During the era of SKA2---with its dramatic sensitivity improvements on SKA1---we will be able to efficiently survey for novae across the Milky Way and explore a full sample of novae in the Magellanic Clouds (Figure~\ref{sphere}). Furthermore, with the improved angular resolution and sensitivity of SKA2, we will be able to resolve the ejecta of novae much earlier in outburst and at even greater distances.

\section{Tackling the big science questions in the era of SKA}

First, we would like to test the theory of mass accretion onto white dwarfs and nova explosions by measuring the ejected mass and explosion energetics for a substantial sample of novae. This will be made possible in the era of SKA because we will have a sample of radio-observed novae with well understood selection effects (and selection effects that should mostly be a simple function of distance and ejected mass), enabling us to model novae as populations. High quality radio light curves complemented by observations at other wavelengths will allow us to finally measure the distribution of ejecta masses and energetics, and compare these with predictions from cataclysmic variable and nova theory. For example, theory makes predictions for the relative numbers of novae with low ejecta mass and high ejecta mass (the massive white dwarfs and high accretion rates that produce novae with low ejecta mass should be rare by number but have short recurrence time, and so should be relatively over-represented in the population), and this prediction will be tested with SKA-detected nova samples.

Second, exquisite multi-frequency light curves from the SKA will enable us to determine the mechanisms of mass ejection in novae, exploring why novae appear so diverse and how the mass-loss mechanism shapes nova ejecta. Novae are famous for their complexity, often showing plateaus, dips, and multiple peaks in their optical light curves \citep{2010AJ....140...34S}, and recent data are showing that their behaviour at radio wavelengths is similarly rich---implying multiple episodes and processes of mass ejection that can act for months. Physically, mass loss in novae may be powered by the impulsive thermonuclear runaway, prolonged optically thick winds, common-envelope like interaction with the binary, jets, or ablation of the companion star/accretion disk. Radio light curves and imaging are powerful tools for disentangling these possibilities, by tracing the bulk of the ejected mass in a simple and straightforward way.

Finally, radio observations of both novae and symbiotics will allow us to map the density and distribution of circumbinary material, enabling a direct comparison between observations of Type Ia supernovae and these proposed progenitor systems. Radio surveys for symbiotic stars will reveal the prevalence of nuclear-burning white dwarfs approaching the Chandrasekhar mass.

In all of the above cases, the real advance of the SKA is that it will give us complete and well-understood samples of novae and symbiotics, so that the field can be transformed from a series of case studies to an understanding of these systems as populations. We will finally be able to answer questions such as: What is the dominant mechanism of mass loss in novae? How aspherical is the typical wind surrounding a symbiotic star? How well do novae adhere to theoretical expectations? Can white dwarfs really grow in mass by accreting non-degenerate matter?

\bibliographystyle{apj_short_etal}

\bibliography{thermalnovae}

\begin{thebibliography}{54}
\expandafter\ifx\csname natexlab\endcsname\relax\def\natexlab#1{#1}\fi

\bibitem[{{Abdo} {et~al.}(2010){Abdo}, {Ackermann}, {Ajello}, {Atwood},
  {Baldini}, {Ballet}, {Barbiellini}, {Bastieri}, {Bechtol}, {Bellazzini}, \&
  et~al.}]{2010Sci...329..817A}
{Abdo}, A.~A. \etal . 2010, Science, 329, 817

\bibitem[{{Ackermann} {et~al.}(2014){Ackermann}, {Ajello}, {Albert}, {Baldini},
  {Ballet}, {Barbiellini}, {Bastieri}, {Bellazzini}, {Bissaldi}, {Blandford},
  {Bloom}, {Bottacini}, {Brandt}, {Bregeon}, {Bruel}, {Buehler}, {Buson},
  {Caliandro}, {Cameron}, {Caragiulo}, {Caraveo}, {Cavazzuti}, {Charles},
  {Chekhtman}, {Cheung}, {Chiang}, {Chiaro}, {Ciprini}, {Claus},
  {Cohen-Tanugi}, {Conrad}, {Corbel}, {D'Ammando}, {de Angelis}, {den Hartog},
  {de Palma}, {Dermer}, {Desiante}, {Digel}, {Di Venere}, {do Couto e Silva},
  {Donato}, {Drell}, {Drlica-Wagner}, {Favuzzi}, {Ferrara}, {Focke},
  {Franckowiak}, {Fuhrmann}, {Fukazawa}, {Fusco}, {Gargano}, {Gasparrini},
  {Germani}, {Giglietto}, {Giordano}, {Giroletti}, {Glanzman}, {Godfrey},
  {Grenier}, {Grove}, {Guiriec}, {Hadasch}, {Harding}, {Hayashida}, {Hays},
  {Hewitt}, {Hill}, {Hou}, {Jean}, {Jogler}, {J{\'o}hannesson}, {Johnson},
  {Johnson}, {Kerr}, {Kn{\"o}dlseder}, {Kuss}, {Larsson}, {Latronico},
  {Lemoine-Goumard}, {Longo}, {Loparco}, {Lott}, {Lovellette}, {Lubrano},
  {Manfreda}, {Martin}, {Massaro}, {Mayer}, {Mazziotta}, {McEnery},
  {Michelson}, {Mitthumsiri}, {Mizuno}, {Monzani}, {Morselli}, {Moskalenko},
  {Murgia}, {Nemmen}, {Nuss}, {Ohsugi}, {Omodei}, {Orienti}, {Orlando},
  {Ormes}, {Paneque}, {Panetta}, {Perkins}, {Pesce-Rollins}, {Piron}, {Pivato},
  {Porter}, {Rain{\`o}}, {Rando}, {Razzano}, {Razzaque}, {Reimer}, {Reimer},
  {Reposeur}, {Saz Parkinson}, {Schaal}, {Schulz}, {Sgr{\`o}}, {Siskind},
  {Spandre}, {Spinelli}, {Stawarz}, {Suson}, {Takahashi}, {Tanaka}, {Thayer},
  {Thayer}, {Thompson}, {Tibaldo}, {Tinivella}, {Torres}, {Tosti}, {Troja},
  {Uchiyama}, {Vianello}, {Winer}, {Wolff}, {Wood}, {Wood}, {Wood},
  {Charbonnel}, {Corbet}, {De Gennaro Aquino}, {Edlin}, {Mason}, {Schwarz},
  {Shore}, {Starrfield}, {Teyssier}, \& {Fermi-LAT
  Collaboration}}]{2014Sci...345..554A}
{Ackermann}, M. \etal . 2014, Science, 345, 554

\bibitem[{{Bird} {et~al.}(2007){Bird}, {Malizia}, {Bazzano}, {Barlow},
  {Bassani}, {Hill}, {B{\'e}langer}, {Capitanio}, {Clark}, {Dean}, {Fiocchi},
  {G{\"o}tz}, {Lebrun}, {Molina}, {Produit}, {Renaud}, {Sguera}, {Stephen},
  {Terrier}, {Ubertini}, {Walter}, {Winkler}, \&
  {Zurita}}]{2007ApJS..170..175B}
{Bird}, A.~J. \etal . 2007, \apjs, 170, 175

\bibitem[{{Bode} \& {Evans}(2008)}]{2008clno.book.....B}
{Bode}, M.~F. \& {Evans}, A., eds. 2008, Classical Novae, 2nd Edition
  (Cambridge: Cambridge University Press)

\bibitem[{{Bode} {et~al.}(2006){Bode}, {O'Brien}, {Osborne}, {Page},
  {Senziani}, {Skinner}, {Starrfield}, {Ness}, {Drake}, {Schwarz}, {Beardmore},
  {Darnley}, {Eyres}, {Evans}, {Gehrels}, {Goad}, {Jean}, {Krautter}, \&
  {Novara}}]{2006ApJ...652..629B}
{Bode}, M.~F. \etal . 2006, \apj, 652, 629

\bibitem[{{Braun}(2013)}]{Braun2013}
{Braun}, R. 2013, "SKA1 Imaging Science Performance", Document number
  SKA-TEL-SKO-DD-XXX Revision A Draft 2

\bibitem[{{Brocksopp} {et~al.}(2004){Brocksopp}, {Sokoloski}, {Kaiser},
  {Richards}, {Muxlow}, \& {Seymour}}]{2004MNRAS.347..430B}
{Brocksopp}, C. \etal . 2004, \mnras, 347, 430

\bibitem[{{Chomiuk} {et~al.}(2012){Chomiuk}, {Krauss}, {Rupen}, {Nelson},
  {Roy}, {Sokoloski}, {Mukai}, {Munari}, {Mioduszewski}, {Weston}, {O'Brien},
  {Eyres}, \& {Bode}}]{2012ApJ...761..173C}
{Chomiuk}, L. \etal . 2012, \apj, 761, 173

\bibitem[{{Chomiuk} {et~al.}(2014{\natexlab{a}}){Chomiuk}, {Linford}, {Yang},
  {O'Brien}, {Paragi}, {Mioduszewski}, {Beswick}, {Cheung}, {Mukai}, {Nelson},
  {Ribeiro}, {Rupen}, {Sokoloski}, {Weston}, {Zheng}, {Bode}, {Eyres}, {Roy},
  \& {Taylor}}]{2014Natur.514..339C}
---. 2014{\natexlab{a}}, \nat, 514, 339

\bibitem[{{Chomiuk} {et~al.}(2014{\natexlab{b}}){Chomiuk}, {Nelson}, {Mukai},
  {Sokoloski}, {Rupen}, {Page}, {Osborne}, {Kuulkers}, {Mioduszewski}, {Roy},
  {Weston}, \& {Krauss}}]{2014ApJ...788..130C}
---. 2014{\natexlab{b}}, \apj, 788, 130

\bibitem[{{Condon} \& {Ransom}(2010)}]{CondonRansom}
{Condon}, J.~J. \& {Ransom}, S.~M. 2010,
  {http://www.cv.nrao.edu/course/astr534/ERA.shtml, Essential Radio Astronomy}

\bibitem[{{Crocker} {et~al.}(2001){Crocker}, {Davis}, {Eyres}, {Bode},
  {Taylor}, {Skopal}, \& {Kenny}}]{2001MNRAS.326..781C}
{Crocker}, M.~M. \etal . 2001, \mnras, 326, 781

\bibitem[{{Dewdney} {et~al.}(2013){Dewdney}, {Turner}, {Millenaar}, {McCool},
  {Lazio}, \& {Cornwell}}]{Dewdney2013}
{Dewdney}, P. \etal . 2013, "SKA1 System Baseline Design", Document number
  SKA-TEL-SKO-DD-001 Revision 1

\bibitem[{{Di Stefano}(2010)}]{2010ApJ...719..474D}
{Di Stefano}, R. 2010, \apj, 719, 474

\bibitem[{{Dougherty} {et~al.}(1995){Dougherty}, {Bode}, {Lloyd}, {Davis}, \&
  {Eyres}}]{1995MNRAS.272..843D}
{Dougherty}, S.~M. \etal . 1995, \mnras, 272, 843

\bibitem[{{Eyres} {et~al.}(1996){Eyres}, {Davis}, \&
  {Bode}}]{1996MNRAS.279..249E}
{Eyres}, S.~P.~S. \etal . 1996, \mnras, 279, 249

\bibitem[{{Eyres} {et~al.}(2009){Eyres}, {O'Brien}, {Beswick}, {Muxlow},
  {Anupama}, {Kantharia}, {Bode}, {Gawro{\'n}ski}, {Feiler}, {Evans},
  {Rushton}, {Davis}, {Prabhu}, {Porcas}, \& {Hassall}}]{2009MNRAS.395.1533E}
---. 2009, \mnras, 395, 1533

\bibitem[{{Fujimoto}(1982)}]{1982ApJ...257..767F}
{Fujimoto}, M.~Y. 1982, \apj, 257, 767

\bibitem[{{Gordon} \& {Sorochenko}(2002)}]{2002ASSL..282.....G}
{Gordon}, M.~A. \& {Sorochenko}, R.~L., eds. 2002, Astrophysics and Space
  Science Library, Vol. 282, {Radio Recombination Lines. Their Physics and
  Astronomical Applications}

\bibitem[{{Hillman} {et~al.}(2014){Hillman}, {Prialnik}, {Kovetz}, {Shara}, \&
  {Neill}}]{2014MNRAS.437.1962H}
{Hillman}, Y. \etal . 2014, \mnras, 437, 1962

\bibitem[{{Hjellming} \& {Wade}(1970)}]{1970ApJ...162L...1H}
{Hjellming}, R.~M. \& {Wade}, C.~M. 1970, \apjl, 162, L1

\bibitem[{{Hjellming} {et~al.}(1979){Hjellming}, {Wade}, {Vandenberg}, \&
  {Newell}}]{1979AJ.....84.1619H}
{Hjellming}, R.~M. \etal . 1979, \aj, 84, 1619

\bibitem[{{Hounsell} {et~al.}(2010){Hounsell}, {Bode}, {Hick}, {Buffington},
  {Jackson}, {Clover}, {Shafter}, {Darnley}, {Mawson}, {Steele}, {Evans},
  {Eyres}, \& {O'Brien}}]{2010ApJ...724..480H}
{Hounsell}, R. \etal . 2010, \apj, 724, 480

\bibitem[{{Karovska} {et~al.}(2010){Karovska}, {Gaetz}, {Carilli}, {Hack},
  {Raymond}, \& {Lee}}]{2010ApJ...710L.132K}
{Karovska}, M. \etal . 2010, \apjl, 710, L132

\bibitem[{{Kennea} {et~al.}(2009){Kennea}, {Mukai}, {Sokoloski}, {Luna},
  {Tueller}, {Markwardt}, \& {Burrows}}]{2009ApJ...701.1992K}
{Kennea}, J.~A. \etal . 2009, \apj, 701, 1992

\bibitem[{{Krauss} {et~al.}(2011){Krauss}, {Chomiuk}, {Rupen}, {Roy},
  {Mioduszewski}, {Sokoloski}, {Nelson}, {Mukai}, {Bode}, {Eyres}, \&
  {O'Brien}}]{2011ApJ...739L...6K}
{Krauss}, M.~I. \etal . 2011, \apjl, 739, L6

\bibitem[{{Kwok}(1983)}]{1983MNRAS.202.1149K}
{Kwok}, S. 1983, \mnras, 202, 1149

\bibitem[{{Luna} \& {Sokoloski}(2007)}]{2007ApJ...671..741L}
{Luna}, G.~J.~M. \& {Sokoloski}, J.~L. 2007, \apj, 671, 741

\bibitem[{{Luna} {et~al.}(2013){Luna}, {Sokoloski}, {Mukai}, \&
  {Nelson}}]{2013A&A...559A...6L}
{Luna}, G.~J.~M. \etal . 2013, \aap, 559, A6

\bibitem[{{Munari} {et~al.}(2013){Munari}, {Dallaporta}, {Castellani},
  {Valisa}, {Frigo}, {Chomiuk}, \& {Ribeiro}}]{2013MNRAS.435..771M}
{Munari}, U. \etal . 2013, \mnras, 435, 771

\bibitem[{{Munari} \& {Renzini}(1992)}]{1992ApJ...397L..87M}
{Munari}, U. \& {Renzini}, A. 1992, \apjl, 397, L87

\bibitem[{{Nelson} {et~al.}(2014){Nelson}, {Chomiuk}, {Roy}, {Sokoloski},
  {Mukai}, {Krauss}, {Mioduszewski}, {Rupen}, \&
  {Weston}}]{2014ApJ...785...78N}
{Nelson}, T. \etal . 2014, \apj, 785, 78

\bibitem[{{Nelson} {et~al.}(2012){Nelson}, {Donato}, {Mukai}, {Sokoloski}, \&
  {Chomiuk}}]{2012ApJ...748...43N}
---. 2012, \apj, 748, 43

\bibitem[{{Nichols} {et~al.}(2007){Nichols}, {DePasquale}, {Kellogg},
  {Anderson}, {Sokoloski}, \& {Pedelty}}]{2007ApJ...660..651N}
{Nichols}, J.~S. \etal . 2007, \apj, 660, 651

\bibitem[{{O'Brien} {et~al.}(2006){O'Brien}, {Bode}, {Porcas}, {Muxlow},
  {Eyres}, {Beswick}, {Garrington}, {Davis}, \& {Evans}}]{2006Natur.442..279O}
{O'Brien}, T.~J. \etal . 2006, \nat, 442, 279

\bibitem[{{O'Brien} {et~al.}(2012){O'Brien}, {Yang}, {Paragi}, {Chomiuk},
  {Cheung}, {Nelson}, {Rupen}, {Taylor}, {Sokoloski}, {Mukai}, {Weston}, {Roy},
  {Mioduszewski}, {Bode}, \& {Eyres}}]{2012ATel.4408....1O}
---. 2012, The Astronomer's Telegram, 4408, 1

\bibitem[{{Osborne} {et~al.}(2011){Osborne}, {Page}, {Beardmore}, {Bode},
  {Goad}, {O'Brien}, {Starrfield}, {Rauch}, {Ness}, {Krautter}, {Schwarz},
  {Burrows}, {Gehrels}, {Drake}, {Evans}, \& {Eyres}}]{2011ApJ...727..124O}
{Osborne}, J.~P. \etal . 2011, \apj, 727, 124

\bibitem[{{Roy} {et~al.}(2012){Roy}, {Chomiuk}, {Sokoloski}, {Weston}, {Rupen},
  {Johnson}, {Krauss}, {Nelson}, {Mukai}, {Mioduszewski}, {Bode}, {Eyres}, \&
  {O'Brien}}]{2012BASI...40..293R}
{Roy}, N. \etal . 2012, Bulletin of the Astronomical Society of India, 40, 293

\bibitem[{{Rupen} {et~al.}(2008){Rupen}, {Mioduszewski}, \&
  {Sokoloski}}]{2008ApJ...688..559R}
{Rupen}, M.~P. \etal . 2008, \apj, 688, 559

\bibitem[{{Seaquist} \& {Bode}(2008)}]{2008CNe.SeaquistBode}
{Seaquist}, E.~R. \& {Bode}, M.~F. 2008, in Classical Novae, 2nd Edition, ed.
  M.~F. {Bode} \& A.~{Evans} (Cambridge: Cambridge University Press), 141--166

\bibitem[{{Seaquist} {et~al.}(1989){Seaquist}, {Bode}, {Frail}, {Roberts},
  {Evans}, \& {Albinson}}]{1989ApJ...344..805S}
{Seaquist}, E.~R. \etal . 1989, \apj, 344, 805

\bibitem[{{Seaquist} {et~al.}(1980){Seaquist}, {Duric}, {Israel}, {Spoelstra},
  {Ulich}, \& {Gregory}}]{1980AJ.....85..283S}
---. 1980, \aj, 85, 283

\bibitem[{{Seaquist} {et~al.}(1993){Seaquist}, {Krogulec}, \&
  {Taylor}}]{1993ApJ...410..260S}
---. 1993, \apj, 410, 260

\bibitem[{{Seaquist} \& {Palimaka}(1977)}]{1977ApJ...217..781S}
{Seaquist}, E.~R. \& {Palimaka}, J. 1977, \apj, 217, 781

\bibitem[{{Seaquist} \& {Taylor}(1990)}]{1990ApJ...349..313S}
{Seaquist}, E.~R. \& {Taylor}, A.~R. 1990, \apj, 349, 313

\bibitem[{{Seaquist} {et~al.}(1984){Seaquist}, {Taylor}, \&
  {Button}}]{1984ApJ...284..202S}
{Seaquist}, E.~R. \etal . 1984, \apj, 284, 202

\bibitem[{{Shafter}(1997)}]{1997ApJ...487..226S}
{Shafter}, A.~W. 1997, \apj, 487, 226

\bibitem[{{Sokoloski} {et~al.}(2006){Sokoloski}, {Luna}, {Mukai}, \&
  {Kenyon}}]{2006Natur.442..276S}
{Sokoloski}, J.~L. \etal . 2006, \nat, 442, 276

\bibitem[{{Sokoloski} {et~al.}(2008){Sokoloski}, {Rupen}, \&
  {Mioduszewski}}]{2008ApJ...685L.137S}
---. 2008, \apjl, 685, L137

\bibitem[{{Strope} {et~al.}(2010){Strope}, {Schaefer}, \&
  {Henden}}]{2010AJ....140...34S}
{Strope}, R.~J. \etal . 2010, \aj, 140, 34

\bibitem[{{Tang} {et~al.}(2014){Tang}, {Bildsten}, {Wolf}, {Li}, {Kong}, {Cao},
  {Cenko}, {De Cia}, {Kasliwal}, {Kulkarni}, {Laher}, {Masci}, {Nugent},
  {Perley}, {Prince}, \& {Surace}}]{2014ApJ...786...61T}
{Tang}, S. \etal . 2014, \apj, 786, 61

\bibitem[{{Taylor} {et~al.}(1988){Taylor}, {Hjellming}, {Seaquist}, \&
  {Gehrz}}]{1988Natur.335..235T}
{Taylor}, A.~R. \etal . 1988, \nat, 335, 235

\bibitem[{{van den Heuvel} {et~al.}(1992){van den Heuvel}, {Bhattacharya},
  {Nomoto}, \& {Rappaport}}]{1992A&A...262...97V}
{van den Heuvel}, E.~P.~J. \etal . 1992, \aap, 262, 97

\bibitem[{{Yaron} {et~al.}(2005){Yaron}, {Prialnik}, {Shara}, \&
  {Kovetz}}]{2005ApJ...623..398Y}
{Yaron}, O. \etal . 2005, \apj, 623, 398

\end{thebibliography}

\end{document}